\documentclass[aps,prl,twocolumn,groupedaddress]{revtex4}

\usepackage{graphicx}
\usepackage{color}
\newcommand{\nc}{\newcommand}
\nc{\black}{\color{black}}
\nc{\red}{\color{red}}
\nc{\blue}{\color{blue}}
\nc{\green}{\color{green}}
\nc{\yellow}{\color{yellow}}
\nc{\orange}{\color{orange}}
\nc{\violet}{\color{violet}}
\nc{\magenta}{\color{magenta}}
\nc{\grey}{\color{grey}}

\begin{document}
\title{Real-space observation of current-driven domain wall motion %
	   \linebreak in submicron magnetic wires}

\author{A.~Yamaguchi}
\affiliation{Graduate School of Engineering Science, Osaka University, Toyonaka 560-8531, Japan}

\author{T.~Ono}
\affiliation{Graduate School of Engineering Science, Osaka University, Toyonaka 560-8531, Japan}

\author{S.~Nasu}
\affiliation{Graduate School of Engineering Science, Osaka University, Toyonaka 560-8531, Japan}

\author{K.~Miyake}
\affiliation{Institute for Chemical Research, Kyoto University, Uji 611-0011, Japan}

\author{K.~Mibu}
\affiliation{Research Center for Low Temperature and Materials Sciences, Kyoto University, Uji 611-0011, Japan}

\author{T.~Shinjo}
\affiliation{International Institute for Advanced Studies, Soraku-gun, 619-0225, Japan}

\date{\today}

\begin{abstract}
We report direct observation of current-driven magnetic domain wall (DW)
displacement by using a well-defined single DW in a micro-fabricated
magnetic wire with submicron width. Magnetic force microscopy visualizes
that a single DW introduced in a wire is displaced back and forth by
positive and negative pulsed-current, respectively. The direct observation gives quantitative information on the DW displacement as a function of the intensity and the duration of the pulsed-current. The result is discussed in terms of the spin-transfer mechanism.

\end{abstract}
\maketitle

In general, ferromagnets are composed of magnetic domains, within each of 
which magnetic moments align. The directions of magnetization of neighboring 
domains are not parallel. As a result, there is a magnetic DW between 
neighboring domains. The direction of moments gradually changes in a DW. 
What will happen if an electric current flows through a DW? Since the spin 
direction of conduction electrons changes when the electrons cross the DW, 
spin transfer from electrons into the DW occurs and torque is exerted on the 
DW. In consequence, the electric current can displace the 
DW\cite{Berger84,Berger92,Tatara,Waintal}. This current-driven DW motion has 
been confirmed by experiments on magnetic thin films and magnetic wires\cite{Berger85,Berger88,Gan,Koo,Grollier,Vernier}. However, quantitative experiment on a single DW in a magnetic wire for getting deeper insights into the physical mechanisms of this effect is still lacking. Our real-space observation by 
MFM gives the quantitative information: DW displacement as a function of the 
intensity and the duration of the pulsed-current. It is found that the DW displacement is proportional to the pulse duration and the DW velocity increases with the current density.

\begin{figure}[htbp]
\caption{Schematic illustration of a top view of the sample. 
One end of the L-shaped wire is connected to a diamond-shaped pad which 
acts as a domain wall (DW) injector, and the other end is sharply pointed 
to prevent a nucleation of a DW from this end. The wire has four electrodes 
made of Cu. MFM observations were performed for the hatched area at room 
temperature.}
\end{figure}

\begin{figure}[htbp]
\caption{(a) MFM image after the introduction of a DW. 
DW is imaged as a bright contrast, which corresponds to the stray 
field from positive magnetic charge. 
(b) Schematic illustration of a magnetic domain 
structure inferred from the MFM image. DW has a head-to-head structure. 
(c) Result of micromagnetics simulation (vortex DW). 
(d) Result of micromagnetics simulation (transverse DW). 
(e) MFM image calculated from the magnetic structure shown in Fig. 2(c). 
(f) MFM image calculated from the magnetic structure shown in Fig. 2(d). 
(g) Magnified MFM image of a DW. 
(h) MFM image after an application of a pulsed-current from left to right. 
The current density and pulse duration are $1.2 \times 10^{12}~$A/m$^{2}$ 
and 5~$\mu$s, respectively. DW is displaced from right to left by the 
pulsed-current. 
(i) MFM image after an application of a pulsed-current from right to left. 
The current density and pulse duration are $1.2 \times 10^{12}~$A/m$^{2}$ 
and $5~\mu$s, respectively. DW is displaced from left to right by the 
pulsed-current.}
\end{figure} 

\begin{figure}[htbp]
\caption{(a)-(k) Successive MFM images with one pulse 
applied between each consecutive image. The current density and the pulse 
duration were $1.2 \times 10^{12}~$A/m$^{2}$ and 0.5~$\mu$s, respectively. 
Note that a tail-to-tail DW is introduced, which is imaged as a dark 
contrast.}
\end{figure} 

We designed a special L-shaped magnetic wire with a round corner as 
schematically illustrated in Fig. 1. One end of the L-shaped magnetic wire 
is connected to a diamond-shaped pad which acts as a DW 
injector\cite{Shigeto}, and the other end is sharply pointed to prevent a 
nucleation of a DW from this end\cite{Schrefl}. L-shaped magnetic wires of 
10 nm-thick Ni$_{81}$Fe$_{19}$ were fabricated onto thermally oxidized Si 
substrates by means of an e-beam lithography and a lift-off method. The width 
of the wire is 240 nm. The wire has four electrodes made of nonmagnetic 
material, 20 nm-thick Cu, for electrical transport measurements. MFM 
observations were performed for the hatched area in Fig. 1 at room 
temperature. CoPtCr low moment probes were used in order to minimize the 
influence of the stray field from the probe on the DW in the wire.

Because of the special shape of the wire, a single DW can be introduced from 
the diamond-shaped pad and it stops in the vicinity of the round corner when a 
magnetic field is applied along the wire axis connected to the 
pad\cite{Shigeto,Allwood_A}. In order to introduce a DW at the position a 
little bit away from the corner, the direction of the external magnetic field 
was set 26 degrees from the wire axis in the substrate plane as shown in 
Fig.1. In the initial stage, a magnetic field of +1 kOe was applied in order 
to align the magnetization in one direction along the wire. Then, a single DW 
was introduced by applying a magnetic field of -175 Oe. After that, the MFM 
observations were carried out in the absence of a magnetic field. The 
existence of the single DW in the vicinity of the corner was confirmed as 
shown in Fig. 2(a). The DW is imaged as a bright contrast, which corresponds 
to the stray field from positive magnetic charge. In this case, a head-to-head 
DW is realized as schematically illustrated in Fig. 2(b). The position and the 
shape of the DW were unchanged after several MFM scans, indicating that the DW 
was pinned by a local structural defect as reported by 
Nakatani $et$ $al.$\cite{Nakatani} and that a stray field from the probe 
was too small to change the magnetic structure and position of the DW. 

To clarify the magnetic structure of the head-to-head DW, micromagnetics 
simulations were performed by using micromagnetics simulator (OOMMF) from 
NIST\cite{NIST}. The parameters used for the calculation were a unit cell size 
of 5 nm $\times$ 5 nm with a constant thickness of 10 nm, a magnetization of 
1.08 T, and a damping constant of $\alpha = 0.1$. The size of the calculated 
model was the same as the sample for the experiment except for the length of 
the wire. Two types of DWs, vortex and transverse DW, were obtained as a 
stable state in the absence of a magnetic field by changing the initial 
magnetization configuration. Figures 2(c) and 2(d) show the results of the 
micromagnetics simulations for the vortex and the transverse DW, respectively. 
Figures 2(e) and 2(f) show the MFM images calculated from the magnetic 
structures\cite{Saito} shown in Figs. 2(c) and 2(d), respectively. By 
comparing the calculated MFM images with the observed high-resolution MFM 
image of the DW (Fig. 2(g)), it is concluded that the DW is the vortex type. 

After the observation of Fig. 2(a), a pulsed-current was applied through the 
wire in the absence of a magnetic field. The current density and the pulse 
duration were $1.2 \times 10^{12}~$A/m$^{2}$ and 5~$\mu$s, 
respectively, and the 
rise and fall times were shorter than 15 ns. Figure 2(h) shows the MFM image 
after an application of the pulsed-current from left to right. The DW, which 
had been in the vicinity of the corner (Fig. 2(a)), was displaced from right 
to left by the application of the pulsed-current. Thus, the direction of the 
DW motion is opposite to the current direction. Furthermore, the direction of 
the DW motion can be reversed by switching the current polarity as shown in 
Fig. 2(i). These results are consistent with the spin transfer 
mechanism\cite{Berger84,Berger92,Tatara,Waintal}. The critical current density
$j_{c}$ below which the DW cannot be driven by the current was observed to be 
about $1.0 \times 10^{12}$~A/m$^{2}$.

Figures 3(a)-3(k) are successive MFM images with one pulsed-current applied 
between each consecutive image. The current density and the pulse duration 
were $1.2 \times 10^{12}~$A/m$^{2}$ and 0.5~$\mu$s, 
respectively. Prior to the MFM 
observation, a magnetic field of -1 kOe was applied in order to align the 
magnetization in the direction opposite to that in the previous experiment. 
Then, a tail-to-tail DW was introduced by applying a magnetic field of 
+175 Oe. The introduced DW is imaged as a dark contrast in Fig. 3, which 
indicates that a tail-to-tail DW is formed as schematically illustrated in 
Fig. 3. The direction of the tail-to-tail DW motion is also opposite to the 
current direction. The fact that both head-to-head and tail-to-tail DWs are 
displaced opposite to the current direction clearly indicates that the DW 
motion is not caused by a magnetic field generated by the current (Oersted 
field). Each pulse displaced the DW opposite to the current direction. 
The difference in the displacement for each pulse is possibly due to the 
pinning by randomly located defects. The average displacement per one 
pulse did not depend on the polarity of the pulsed-current. 

We discuss the interpretation of the observed current-driven DW motion. 
The Joule heating by the pulsed-current 
should have some effect on the DW motion because it avtivates the thermal 
process. However, the heating cannot explain the fact that the direction 
of the DW motion is reversed by switching the current polarity. The effect 
of the Oersted field is also ruled out as mentioned above. Hydromagnetic 
DW-drag force associated with the Hall effect is negligible in films thinner 
than 0.1~$\mu$m\cite{Berger78}. Therefore, only the spin transfer 
mechanism\cite{Berger84,Berger92,Tatara,Waintal} can explain our experimental 
results. 

For more quantitative discussion, we investigated the DW displacement as 
a function of the duration and the intensity of the pulsed-current. 
Figure 4(a) shows the average DW displacement per one pulse as a function 
of the pulse duration under a condition of constant current density of 
$1.2 \times 10^{12}~$A/m$^{2}$. The average DW displacement is almost 
proportional 
to the pulse duration, which indicates that the DW has a constant velocity 
of 3.0 m/s and the acceleration of the DW can be neglected in the time domain investigated. Figure 4(b) 
shows the average DW velocity as a function of the current density. The 
average velocity could be determined only in a narrow range of the current 
density from $1.1 \times 10^{12}~$A/m$^{2}$ to $1.3 \times 10^{12}~$A/m$^{2}$. 
Below $1.1 \times 10^{12}~$A/m$^{2}$, the displacement for each pulse was not 
reproducible. Above $1.3 \times 10^{12}~$A/m$^{2}$, the samples were degraded 
by the Joule heating due to the high current density. Although the DW 
velocity was measured only in the small current density range, it was well 
confirmed that the DW velocity increases with the current density, as 
expected from the spin transfer mechanism. 

\begin{figure}[htbp]
\centerline{\includegraphics[width=8cm]{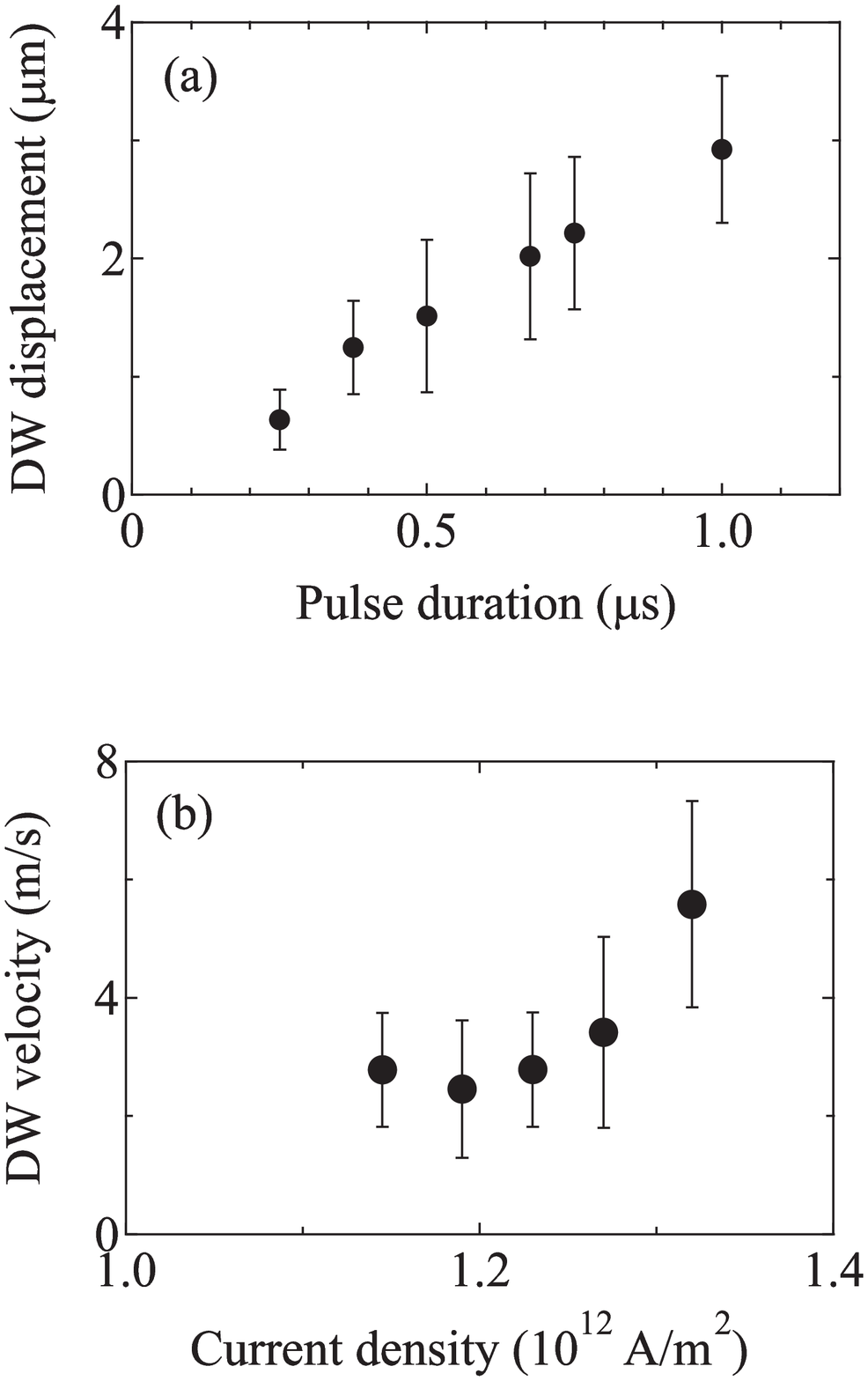}}
\caption{(a) Average DW displacement per one pulse as a function of the 
pulse duration under a condition of constant current density of 
$1.2 \times 10^{12}~$A/m$^{2}$. 
(b) Average DW velocity as a function of the current density.}
\end{figure} 

Since the DW width in the present experiment is much larger 
than the Larmor precession length (several nm), the electron's spin can 
adiabatically follow the direction of the local 
magnetization\cite{Berger84,Berger92,Tatara,Waintal}. As a result, each electron passing 
through the DW flips its spin and gives a quantum $\hbar$ of angular momentum 
to the DW. The displacement of the DW should be proportional 
to the amount of the transferred angular momentum, which is proportional to the duration of the pulsed-current.  Thus, the finding that the DW displacement is proportional 
to the pulse duration supports that the observed DW motion is due to the spin-transfer. Here, we discuss the efficiency of the current-driven DW motion in the present 
experiments. The spin transfer from an electron adds magnetic moment of $2\mu_{B}$ to the DW, 
where $\mu_{B}$ is the Bohr magnetron. Thus, the expected change of magnetic 
moment in the wire by the pulsed-current, $m_{current}$, is calculated as 
$m_{current} = 2 p \mu_{B} j S \Delta t/e$, where $p$ is the spin-polarization 
of the current, $j$ is the current density, 
$S$ is the cross-sectional area of the wire, $\Delta t$ is the pulse 
duration, and $e$ is the electronic charge. On the other hand, the 
change of magnetic moment in the wire by the displacement of the 
DW, $\Delta m$, is calculated as $\Delta m = 2 M_{S} \Delta l S$, 
where $\Delta l$ is the displacement of the DW. Thus, we define 
the efficiency as $\eta = \Delta m/m_{current}$. From the definition, 
the efficiency is zero below $j_{c}$ because the DW does not move 
below $j_{c}$. This means the transferred spin angular momentum 
dissipates into the environment possibly 
through the excitation of local spin waves in the DW. The efficiency 
increases with the current density and $\eta = 0.1$ at 
$j = 1.3 \times 10^{12}~$A/m$^{2}$ if we assume $p = 0.7$\cite{Bass}. 

We have shown the current-driven DW motion for a single DW with a 
well-defined magnetic structure in a submicron magnetic wire, which 
certifies the spintronic device operation\cite{Vers,Allwood_S} by 
this effect. All experimental results are qualitatively consistent with the spin 
transfer mechanism\cite{Berger84,Berger92,Tatara,Waintal}. 
It was found that merely several percent of the transferred angular momentum was used for the displacement of the DW. This is possibly due to the complicated magnetic structure in the vortex DW. Detailed experiments by changing the thickness and the width of the wire are needed to elucidate the origin of the low efficiency.

\begin{acknowledgments}
We would like to thank Y.~Suzuki, S.~Yuasa, H.~Kohno, and 
G.~Tatara for valuable discussions. The present work was partly supported 
by the Ministry of Education, Culture, Sports, Science and Technology of 
Japan (MEXT), through the Grants-in-Aid for COE Research 
(10CE2004 and 12CE2005) and MEXT Special Coordination Funds for 
Promoting Science and Technology 
(Nanospintronics Design and Realization, NDR).
\end{acknowledgments}
 
\end{document}